\documentclass{article}
\usepackage[utf8]{inputenc}
\title{Nonlinear mechanics of colloidal gels:\\creep, fatigue and shear-induced yielding}
\author{Thomas Gibaud$^1$, Thibaut Divoux$^{1,2}$, S\'ebastien Manneville$^1$}
  
\usepackage[dvipsnames]{xcolor}
\usepackage{amssymb}
\usepackage{graphicx}

\date{%
    $^1$Univ Lyon, Ens de Lyon, Univ Claude Bernard, CNRS, Laboratoire de Physique, F-69342 Lyon, France\\%
    $^2$MultiScale Material Science for Energy and Environment, UMI 3466, CNRS-MIT, 77 Massachusetts Avenue, Cambridge, Massachusetts 02139, USA\\[2ex]%
    \today
    }

\begin{document}

\maketitle

\section*{Glossary}

\begin{itemize}

    \item \textbf{Colloids} -- Refers to entities, such as particles or polymers, smaller than a few microns in size and sensitive to thermal agitation. Colloids therefore display Brownian motion when dispersed at low volume fraction in a solvent. They encompass a broad range of entities from proteins and clay particles to synthetic polymeric particles and come in all sorts of geometrical shapes, such as spheres, platelets and rods. Colloids are also subject to a wide variety of interactions, which gives rise to rich state diagrams including fluid and crystalline equilibrium phases as well as out-of-equilibrium states such as gels and glasses.

    \item \textbf{Colloidal gels --} Refers to soft solids, whose structure is composed of attractive colloids that form a space-spanning network. Colloidal gels are in an out-of-equilibrium state and display mainly solid-like properties at rest and under mild deformations, i.e. their elastic modulus is much larger than their viscous modulus in the linear regime. Upon increasing external mechanical constraints, their viscoelastic response becomes nonlinear and generally involves a yield point above which colloidal gels show liquid-like properties (see ``Yielding'' below).
    
    \item \textbf{Creep --} Refers to the motion induced by imposing a constant load to a material, whose response is quantified by the resulting deformation or strain.
    
    \item \textbf{Fatigue --} Refers to the phenomena induced by imposing an oscillatory load of constant amplitude to a material, whose response is quantified via the temporal evolution of its viscoelastic properties.
    
    \item \textbf{Yielding -- } Refers to a solid-to-liquid transition induced by imposing some deformation or stress to a material. The yield stress and yield strain are respectively the critical stress and strain above which a complex material starts to flow or loses its integrity. The yielding transition often involves a complex spatio-temporal scenario, which is still the topic of intense research effort.   

\end{itemize}

\section*{Definition of the Subject}

Colloidal gels are formed through the aggregation of attractive particles, whose size ranges from 10~nm to a few micrometers, suspended in a liquid. Such gels are ubiquitous in everyday life applications, from food products to paints or construction materials, in particular thanks to their ability to easily ``yield'', i.e., to turn from a solid to a liquid under the application of a weak external load. Understanding and controlling the mechanical response of colloidal gels is therefore of prime importance. Depending on the details of the system, however, the resulting gel networks present different microstructural organisations that may lead to widely different mechanical responses. This raises important challenges in fully characterizing yielding and in uncovering the mechanisms of nonlinear response in colloidal gels. In this paper, we distinguish between two classes of colloidal gels showing respectively reversible yielding, where the gel network reforms upon load release, and irreversible yielding, where the network is fully destroyed through fractures and phase separation. This broad, empirical distinction is achieved through rheology and local experiments at a mesoscopic scale, intermediate between the network characteristic size and the sample size. We further discuss how the observables derived from creep and fatigue experiments may be modelled to predict yielding and highlight open questions and future research directions in the domain.

\section{Introduction: from gelation to rupture in colloidal dispersions}

\subsection{General context: the mechanics of colloidal gels}

In mechanics, material rupture refers to a catastrophic series of microscopic events that lead to the macroscopic, irreversible failure of a material under an applied load. Studying the onset of rupture and rupture itself is crucial in fields like earthquake studies and construction materials where understanding and being able to predict the rupture of hard materials is at the core of people's safety \cite{Sornette:2002,Alava:2006,Dixon:2014}. In soft solids such as gels, rupture is less dramatic but just as important.

A colloidal gel forms whenever attractive colloidal particles dispersed in a fluid aggregate into a loose, fragile network that percolates across the sample volume \cite{Larson:1999,Mewis:2012}. Colloidal gels are found in a huge number of practical situations, from pharmaceutical and food products \cite{gibaud2012,Mezzenga:2013} to construction materials like cement \cite{lootens2004,Ioannidou:2016}, as well as ink-jet printing \cite{lewis2002,smay2002,Tan:2018} or flow-cell batteries \cite{youssry2013,Fan:2014,Helal:2016,Narayanan:2017}. Such ubiquity results from the peculiar mechanical response of colloidal gels: while they behave like soft elastic solids at rest, with an elastic modulus in the range of 1~Pa to 10~kPa, colloidal gels are easily made to flow upon application of mild external stresses including shear, compression, gravity or vibration. This solid-to-liquid out-of-equilibrium transition is referred to as the ``yielding transition,'' and a fundamental understanding of the various steps involved in such a transition is of prime importance for applications to material design and industrial processes \cite{Balmforth:2014,Bonn:2017}. For instance, in food science, the texture and yielding properties of edible gels control mouth feel \cite{Jetlema:2016,Wagner:2017} but is also at the heart of food design for people with swallowing problems such as toddlers or elderly people \cite{aguilera2016}. 

Depending on the nature of the particles and due to the wide variety of ways to induce gelation, e.g., through electrostatic interactions, dipole-dipole interactions, pH, or temperature variations, gel networks present a broad variety of structural organizations at the microscopic scale \cite{Nicolai:2013,Mezzenga:2013}. The mechanical properties of colloidal gels are then encoded in the colloidal interactions and the gel structure. In particular, the gel structure is expected to have a significant impact on the way the gel fails under stress. Yet, uncovering the exact link between the interparticle potential, the gel structure, and the gel mechanical properties is tricky and currently constitutes an active research domain. Such a difficulty can be mainly attributed to the fact that colloidal gels are out-of-equilibrium phases with a complex microstructure that is itself pushed into a nonlinear regime due to the applied stress. Rationalizing the link between the gel structure and the failure scenario, therefore, remains a complex exercise that we shall approach here by considering two limiting cases, both in the gel structure and in the rupture mechanisms. Indeed, the main goal of this paper is to apprehend the various yielding processes in colloidal gels through two systems with similar elasticity at rest and yet very distinct rupture mechanisms: a carbon black gel that shows \textit{reversible} fluidization upon yielding (also commonly referred to as shear-rejuvenation) and a sodium caseinate gel that displays \textit{irreversible} brittle failure akin to hard solids (also referred to as brittle-like failure). 

In the rest of this section, we first briefly remind the reader about the most common gelation pathways in attractive colloids and describe the two gel systems that we will use to illustrate our discussion on yielding processes. We then introduce the rheological measurements that are usually performed to assess the nonlinear mechanical response of colloidal gels. In Section~\ref{sec:two_types}, we distinguish experimentally between two types of yielding behaviors in colloidal gels, namely reversible vs.~irreversible yielding, both from a macroscopic point of view and from local approaches. Finally, Section~\ref{sec:discussion} further generalizes the previous observations based on the current literature and highlights open questions in the domain.

\subsection{A brief reminder on colloidal gelation}

\begin{figure}[hpt!]
	\centering
  \includegraphics[width=11cm]{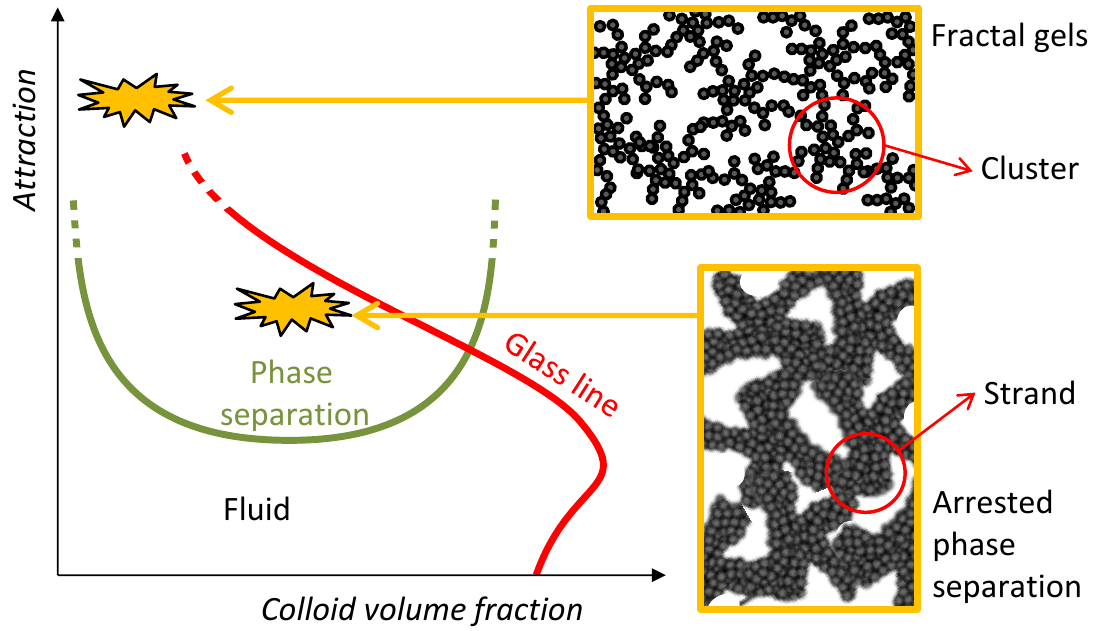}
     \caption{{\bf Schematic state diagram for colloidal particles with short-range attraction}. In the extreme case of very strong interparticle attraction and very low volume fractions, colloids aggregate into fractal gels whereas at intermediate strength of attraction and volume fractions, gels result from the interplay between phase separation (the green line indicates the binodal line) and the glass transition (red line). Those two types of gels have very different structures, characterized by clusters for fractal gels and network strands for arrested phase separation gels.
     }
    \label{fig:statediag}
\end{figure}

As sketched in Fig.~\ref{fig:statediag}, two limit scenarios can be distinguished in the case of colloids with short-range attraction \cite{trappe2004,zaccarelli2007}. In the extreme case of very strong interparticle attraction and very low volume fractions, the regime of so-called ``irreversible aggregation'' is reached, and one observes the formation of soft fractal gels \cite{shih1990,krall1998}. In this regime, the intermediate entities between the particle size and the gel network are fractal clusters, i.e., fractal aggregates of individual colloids. The particles within the clusters are trapped in the potential well of their neighbors so that their motion is constrained, thereby  leading to arrested dynamics and providing elasticity to the network. At intermediate attraction strength and concentrations, the regime of so-called ``arrested phase separation'' is reached \cite{foffi2005,cardinaux:2007,buzzaccaro2007,lu2008}. In this regime, a gel results from the interplay between phase separation and the glass transition line: the dispersion is unstable and phase separates into a bi-continuous network (spinodal structure) composed of a ``gas'' phase, i.e., a dilute phase of colloids, and a ``liquid'' phase, i.e., a dense phase of colloids. During the phase separation process, the concentration of the liquid phase increases and eventually meets the glass line where the liquid network becomes glassy, thus freezing the process of phase separation. Such arrested dynamics leads to a long-lived  network of particles with a spinodal structure \cite{gibaud2009,gibaud2012,gao2015,da2016}. Here, the intermediate entities between the particle size and the gel network are the network strands. Strands are composed of densely packed attractive particles and provide rigidity to the network.  

Although very simplified, Figure~\ref{fig:statediag} illustrates how the gelation kinetic pathway may affect the microstructure of the resulting colloidal gel. Since the microstructure directly impacts the mechanical properties of these out-of-equilibrium systems, controlling the route to gelation offers a way to tune gel properties. A common way to control gelation consists in using particles with an interaction potential that is sensitive to an external stimulus such as pH or temperature \cite{gosal2000}. By increasing the attraction strength, one can continuously go from a fluid state to a gel state. Another common way to control the gel state consists in playing with the shear history experienced by the system during or after gelation \cite{Ovarlez:2013,Koumakis:2015,Helal:2016,Narayanan:2017,Hipp:2019,jamali2020}. By tuning the way a colloidal gel is sheared or mixed, one may vary its terminal structure and mechanical strength without changing the interparticle potential. Two experimental knobs may be controlled: the preshear intensity and the rate of flow cessation. On the one hand, gels reformed after strong shearing and ``instantaneous" flow cessation generally evolve into stronger solids with a relatively homogeneous and fine microstructure, whereas the application of weak shear rates leads to weaker gels with a coarser microstructure \cite{Koumakis:2015}. On the other hand, a preshear of large intensity followed by flow-cessation experiments  performed at different rates allows one to tune continuously the gel microstructure. For instance, it was shown for some fractal gels that the slower the flow cessation, the less connected and the weaker the resulting gel \cite{Helal:2016}. A major difference in the final gel states accessed via continuous quenching of pH or temperature and mechanical shearing lies in the anisotropy and residual stresses associated with the latter \cite{Grenard:2014,Colombo:2017,Sung:2018}.

\subsection{Two illustrations of colloidal gel microstructures}

To apprehend the mechanical properties of colloidal gels, we focus on two systems with distinct gelation pathways and microstructures: (i)~a dispersion of carbon black particles in oil \cite{Trappe:2000} and (ii)~an acid-induced sodium caseinate gel \cite{Bremer:1990,Braga:2006}. 

\begin{figure}
 	\centering
 	\includegraphics[width=.9\columnwidth]{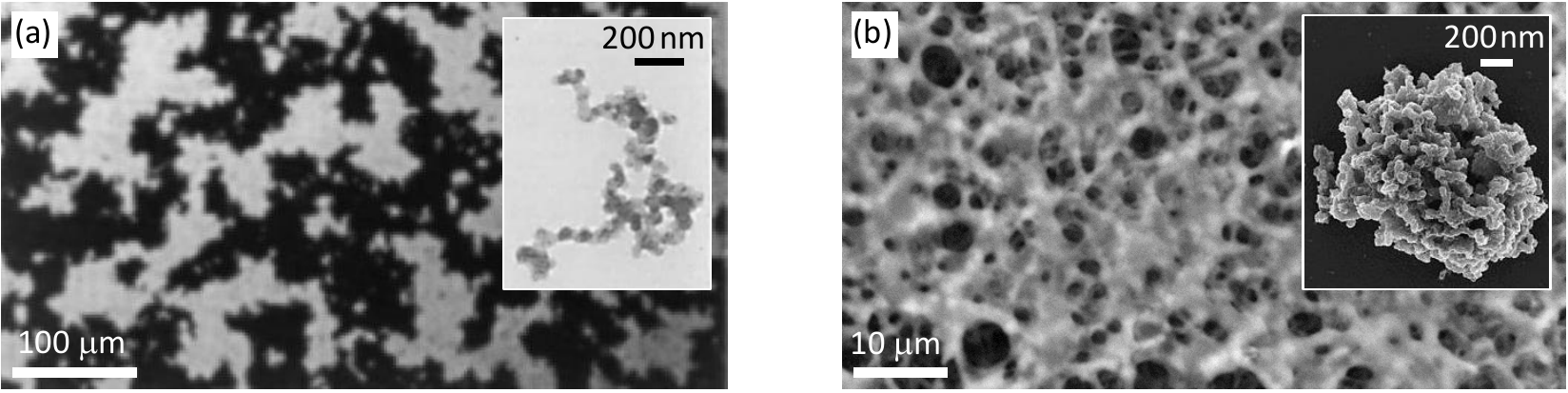}
    \caption{{\bf Microstructures of two colloidal gels.} (a)~Optical microscopy image of a carbon black dispersion at 0.064~vol \% in light mineral oil (adapted from \cite{Trappe:2000} with permission from the American Physical Society). Inset: TEM image of an individual carbon black particle (adapted from \cite{Ehrburger1990} with permission from Elsevier).
    (b)~TEM microscopy image of a sodium caseinate gel at 4~wt \% in water acidified with 1~wt \% GDL (unpublished data). Inset: SEM image of an individual sodium caseinate particle at pH=3.5 (adapted from \cite{Coskun2015} with permission from Elsevier).
    }
     \label{fig:gel}
 \end{figure}

Carbon black gels are composed of carbon black particles dispersed in oil and are a reasonable realization of fractal gels that result from cluster aggregation \cite{Donnet:1993,Trappe:2000}. Carbon black particles are obtained from partial combustion of fuel and are made of unbreakable aggregates of permanently fused nanometric primary particles. These aggregates have a typical diameter of 500~nm. When dispersed in a light mineral oil at a weight concentration of a few percent, carbon black particles form a space-spanning gel network of fractal dimension $d_f\simeq 2.2$ due to attractive Van der Waals interactions of typical strength $U \sim 30k_BT$ \cite{Hartley:1985,Trappe:2007,Richards:2017}. Representative pictures of the carbon black gel microstructure and an individual carbon black particle are shown in Fig.~\ref{fig:gel}(a). The initial gel state is obtained by preshearing the suspension at $+1000$~s$^{-1}$ then at $-1000$~s$^{-1}$ for 20~s each, in order to break up any large aggregates. Abrupt flow cessation quickly leads to the gel formation (in less than a second), which is then left at rest for 100~s before performing any subsequent test. All results concerning carbon black gels presented in this article are based on such a preshear protocol. Note that preshear not only rejuvenates the sample by erasing its previous shear history, it also selects a specific reproducible microstructure with some degree of anisotropy and stored internal stresses that may play a significant role in the subsequent mechanical response \cite{Osuji:2008,Negi:2009,Grenard:2014,Helal:2016}.

Sodium caseinate gels are made of sodium caseinate proteins dispersed in water and can be seen as a close realisation of arrested phase separation. Sodium caseinate is a by-product of casein, which is the main protein component of milk \cite{fox2003}. Casein is present in the form of polydisperse spherical complexes containing casein proteins and colloidal calcium phosphate. Four main casein proteins may be distinguished, namely $\alpha_1$-, $\alpha_2$-, $\beta$- and $\kappa$-casein, with typical weight ratios 0.4:0.08:0.4:0.1 and approximately the same molar mass $M\simeq 2\cdot10^4$~g.mol$^{-1}$ \cite{hadjsadok2008}. The native casein complex dissociates after removal of the colloidal calcium phosphate, yielding a mixture of individual casein proteins referred to as sodium caseinate with an average diameter of 400~nm~\cite{Coskun2015}. Sodium caseinate particles display a mixed interparticle potential: a short-range attraction and a long-range electrostatic repulsion due to the protein surface charges. Gelation can be obtained by turning down the electrostatic repulsion. This can be achieved using the pH as a control parameter thanks to the presence of carboxylate and primary amine groups at the surface of the sodium caseinate, which charge depends on the pH. At pH=7, sodium caseinates are on average positively charged, and an aqueous dispersion of caseinate is stable and translucent. All sodium caseinate gels discussed here are obtained by using glucono-$\delta$-lactone (GDL), a molecule that slowly hydrolyzes into gluconic acid and thus continuously, and homogeneously lowers the pH of the sodium caseinate dispersion from 7 to 3. This GDL-controlled acidification enables the formation of a bulk gel phase around the isoelectric point --the point where surface charges vanish on average-- at pH$_{\rm cas}$=4.6. At pH=pH$_{\rm cas}$, the gelation is kinetically driven by diffusion-limited cluster aggregation (DLCA) \cite{Bremer:1990}. Right below pH$_{\rm cas}$, at pH=3, sodium caseinates have slightly negative surface charges and over-acidification results in the coexistence of a ``gas'' phase of dilute sodium caseinates and a glassy network \cite{Leocmach:2015}, a scenario that is consistent with arrested phase separation \cite{cardinaux:2007}. Figure~\ref{fig:gel}(b) displays an example of transmission electron microscopy (TEM) image of an acid-induced sodium caseinate gel together with a scanning electron microscopy (SEM) image of an individual sodium caseinate particle at pH=3.5. The gel structure is made of strands with a typical width of 1~$\mu$m. All mechanical characterizations presented below are performed on sodium caseinate gels prepared \textit{in situ}, i.e., for which acidification is performed directly in the measurement cell.

\subsection{A first approach of yielding based on oscillatory shear rheometry} 
The two colloidal gels presented above result from very different gelation pathways and thus present very different microstructures. The main question that we wish to address in the following is: how do their nonlinear mechanical responses differ? As a first approach of this question, we now introduce the widespread toolbox of rheology to induce and assess gel yielding.

The mechanical response of soft materials is most commonly probed through shear rheometry, where the sample is confined between a fixed boundary and a rotating tool and the shear stress $\sigma$ and strain $\gamma$ are measured at a macroscopic level, respectively through the torque applied on the rotating tool and its displacement \cite{Macosko:1994}. One usually explores the gel mechanical behavior from the linear elastic regime up to failure thanks to oscillatory shear. More specifically, in a strain-sweep experiment, the rheometer imposes an oscillatory strain $\gamma(t)=\gamma_0\cos(2\pi f t)$ of increasing amplitude $\gamma_0$ at a frequency $f$, and records the corresponding stress response $\sigma(t)$. Linear response theory allows one to define the response function $G^*$ that relates the stress to the strain as $\sigma=G^*\gamma$ in complex notation. $G^*$ is referred to as the complex modulus of the gel and can be decomposed into its real part, the elastic (or storage) modulus $G'$ that indicates how much elastic energy the gel can store, and its imaginary part, the viscous (or loss) modulus $G''$ that indicates how much energy is lost due to dissipation.

 \begin{figure}
 	\centering
 	\includegraphics[width=1\columnwidth]{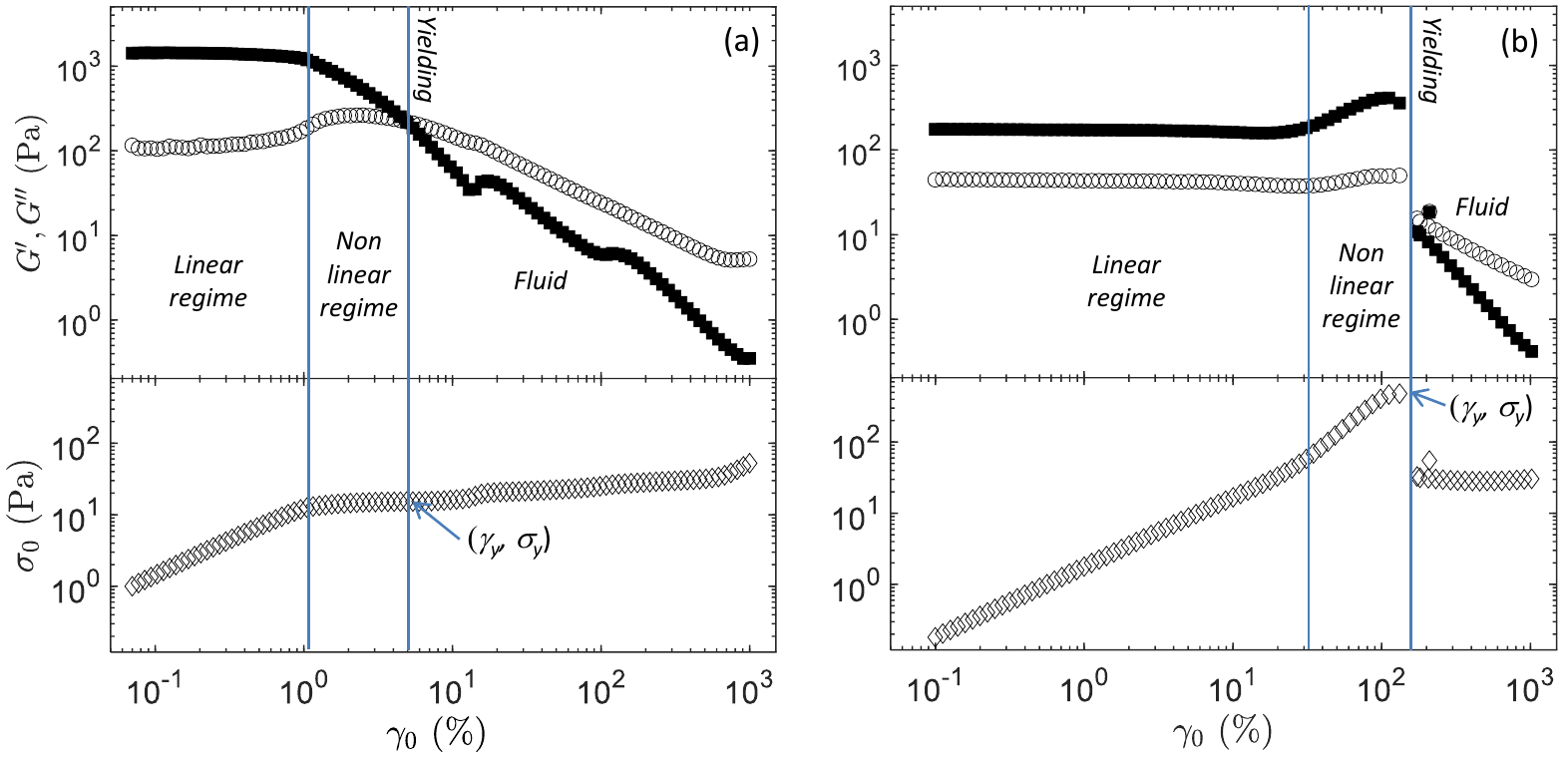}
         \caption{{\bf Assessing yielding from oscillatory shear.} Elastic modulus $G'$ ($\blacksquare$), viscous modulus $G''$ ($\circ$) and stress amplitude $\sigma_0$ ($\lozenge$) as a function of the amplitude $\gamma_0$ of oscillatory strain at frequency $f=1$~Hz in (a)~a carbon black dispersion at 6~wt \% in light mineral oil (adapted from \cite{Perge:2014b} with permission from the Society of Rheology) and (b)~a sodium caseinate gel at 4~wt \% in water acidified with 4~wt \% GDL (unpublished data).}
     \label{fig:rheoyield}
 \end{figure}

Figure~\ref{fig:rheoyield} compares the mechanical properties $G'$ and $G''$ of a carbon black gel [Fig.~\ref{fig:rheoyield}(a)] and those of a sodium caseinate gel [Fig.~\ref{fig:rheoyield}(b)] as a function of the strain amplitude $\gamma_0$. For both systems, the gel response to oscillatory shear can be decomposed into three different regimes depending on $\gamma_0$. At small $\gamma_0$, in the linear viscoelastic regime, the stress amplitude $\sigma_0$ is simply proportional to the strain amplitude $\gamma_0$: the sample mechanical properties are entirely characterized by the elastic modulus $G'$ and loss modulus $G''$. For the gels under study, $G'$ is of the order of 100--1000~Pa and much greater than $G''$, a characteristic of soft solids. Upon increasing $\gamma_0$, the strain response becomes non-linear. Therefore, linear response theory no longer strictly applies, and $G'$ and $G''$ only provide information on the material response at the fundamental frequency $f$. In other words, higher harmonics of the stress $\sigma(t)$ must be taken into account to fully characterize nonlinear viscoelasticity \cite{Hyun:2011}. Eventually, above a critical strain $\gamma_y$ defined as the strain where $G'=G''$, the gel yields and subsequently flows like a liquid: this critical strain $\gamma_y$ corresponds to the {\it yield strain} that can be inferred from large-amplitude oscillatory shear (LAOS) and the corresponding stress is associated with the {\it yield stress} $\sigma_y$. Note that such definitions of the yield strain and stress are obviously protocol-dependent \cite{Bonn:2017} 
and that in the case of the carbon black gel, $G'$ displays two local maxima beyond the yield point, which have been interpreted respectively in terms of inter-cluster and intra-cluster bond-breaking \cite{Koumakis:2011}.

While being qualitatively similar, the behaviors of $G'$ and $G''$ as a function of the strain amplitude hint at different mechanical scenarios for the two gels under study even long before yielding. One may first note that although its elastic modulus at rest is about 10 times larger than that of the sodium caseinate gel, the carbon black gel enters the nonlinear regime and then yields at strain amplitudes that are about 10 times smaller than in the sodium caseinate gel. Thus, the more elastic, fractal gel constituted of particle clusters is more sensitive to external strain than the softer yet more ``resistant'' gel, whose structure is made of strands. Second, in the ``medium-amplitude oscillatory shear'' (MAOS) regime, the carbon black gel shows \textit{strain softening}, i.e., the elastic modulus drops for $\gamma_0\gtrsim 1\%$ and the stress amplitude $\sigma_0$ plateaus as a function of $\gamma_0$, while the sodium caseinate gel displays \textit{strain stiffening}, i.e., $G'$ increases for $\gamma_0\gtrsim 20\%$ and $\sigma_0$ increases faster than linearly with $\gamma_0$. Third, contrary to the carbon black gel that shows a smooth, continuous variation of the various observables at the yielding transition, the sodium caseinate gel displays a strong discontinuity. In particular, the stress abruptly drops at yielding, which is indicative of the sample fracture. Indeed, visual inspection on the sample in a concentric-cylinder geometry reveals that the sodium caseinate gel fractures irreversibly at yielding, as also described in more details in Section~\ref{sec:local} below.

To summarize, the above description of strain sweeps remains only qualitative and does not provide full details on the yielding scenario. Although the rheological response allows estimating the yield stress $\sigma_y$ and the yield strain $\gamma_y$ of the gel, it is insufficient to characterize the gel yielding as it does not provide any time-resolved information. This calls for more refined, time-resolved measurements, both macroscopic and microscopic, to uncover the reasons for the specific variations of $G'$ and $G''$ exemplified in Fig.~\ref{fig:rheoyield}. 

\section{Reversible yielding versus irreversible rupture }
\label{sec:two_types}

 \begin{figure}
 	\centering
 	\includegraphics[width=1\columnwidth]{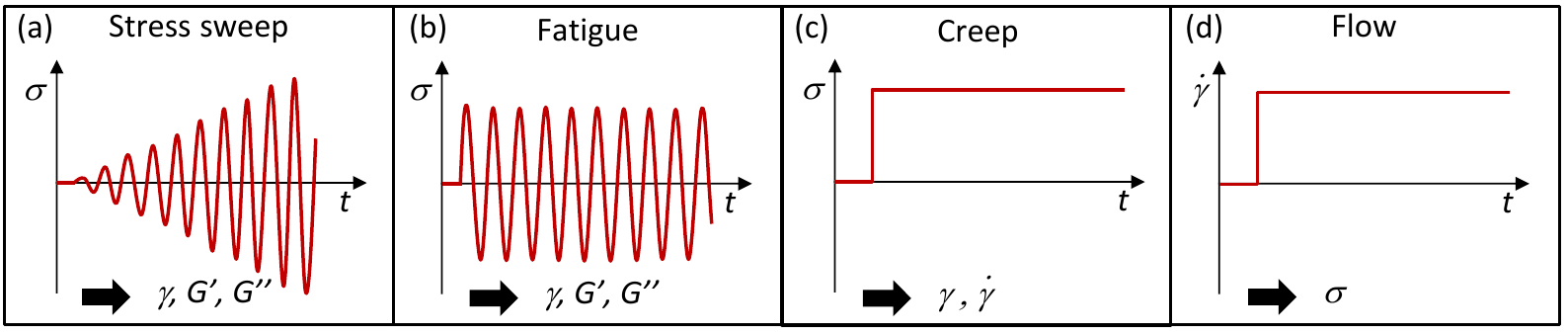}
         \caption{{\bf Rheological tests used to probe yielding in colloidal gels.} (a)~Stress amplitude sweep at fixed frequency. (b)~Fatigue test under an oscillatory shear stress of constant amplitude and frequency. (c)~Creep test under a constant shear stress. (d)~Shear start up test under constant imposed shear rate. The imposed variable in (a)--(c) is the stress $\sigma$. The measured observables are the strain $\gamma$, the elastic modulus $G'$ and the viscous modulus $G''$ in (a,b) and the strain $\gamma$ in (c) or equivalently its time-derivative, the shear rate $\dot\gamma$. In (d), the imposed variable is the shear rate $\dot{\gamma}$ and the measured variable is the stress $\sigma$.}
     \label{fig:test}
 \end{figure}

In this section, we offer a time-resolved analysis of the yielding transition of colloidal gels at the macroscopic level. Indeed, the amplitude sweep protocol used so far to characterize yielding and sketched in Fig.~\ref{fig:test}(a) does not allow one to disentangle the effects of increasing the stress or strain from the temporal evolution of the sample, e.g., due to slow damage accumulation. In fact, there is a major difference between the yielding behaviors of the two systems presented above. Whilst the carbon black gel swiftly reforms and regains elasticity once the external load is released, i.e., yielding and shear-induced fluidization are to some extent \textit{reversible},  the sodium caseinate gel remains liquid-like beyond the yield point, and composed of macroscopic pieces of the original gel network that coexist with the ``gas'' phase of dilute sodium caseinates. The protein gel is permanently broken and the yielding process thus appears as \textit{irreversible}. In amplitude sweep experiments, such a striking difference only shows through the discontinuity reported in Fig.~\ref{fig:gel}(b) and through the fact that one may perform several successive identical sweep tests on the carbon black gels, leading to reproducible results (provided that two consecutive tests are separated by the same preshear step), whereas one needs to prepare a fresh protein gel prior to each experiment.

To better characterize the two types of yielding scenarios and quantify their dynamics, we consider other mechanical tests, namely fatigue and creep tests that are schematized respectively in Fig.~\ref{fig:test}(b) and Fig.~\ref{fig:test}(c). Contrary to stress sweep experiments, the input signal is stationary, which facilitates the analysis of the material response and its time dependence. The primary objective of those tests is to determine the fluidization (or fracture) time $\tau_f$ defined as the time needed for the gel to become fluid as a function of the input parameter. Stress-controlled experiments such as fatigue or creep tests provide a relationship $\tau_f$ vs. $\sigma$, which characterizes the yielding dynamics at the macroscopic scale. We will show that the functional form of $\tau_f(\sigma)$ is representative of the rupture mechanism and allows one to differentiate between reversible and irreversible yielding. We will then briefly turn to local measurements that provide additional insight into the yielding process at the ``mesoscale,'' i.e., at spatial scales intermediate between the macroscopic scale probed by rheometry and the microscopic scale of the colloidal gel network.

\subsection{Carbon black gels as model materials for ``reversible" yielding}

 \begin{figure}
 	\centering
 	\includegraphics[width=0.9\columnwidth]{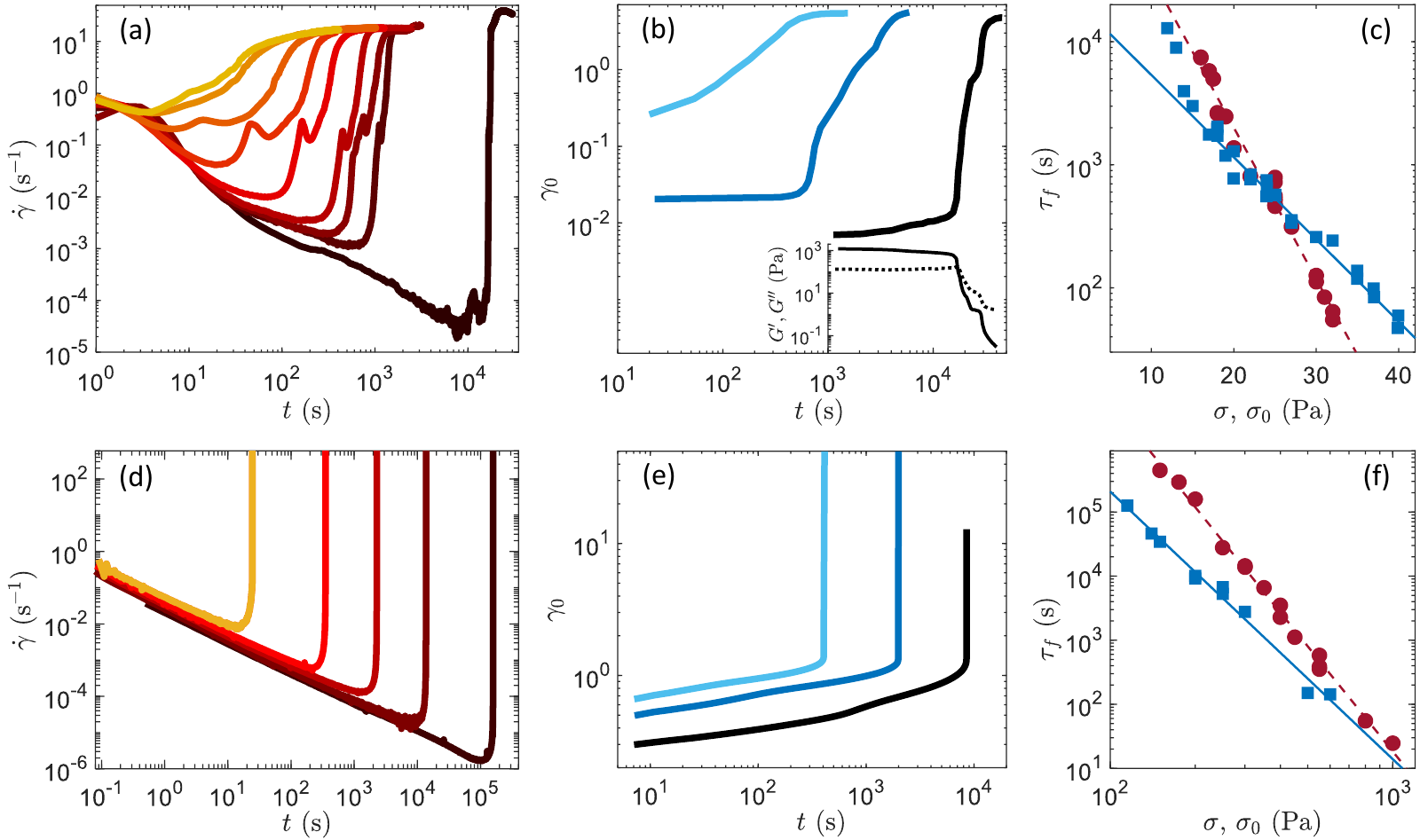}
         \caption{{\bf Comparison between creep and fatigue experiments in carbon black gels and sodium caseinate gels}. {\bf (a)--(c) Delayed yielding in carbon black gels.} (a)~Creep experiments in a carbon black gel at 8~wt~\% in light mineral oil: shear rate responses $\dot\gamma(t)$ for different shear stresses $\sigma$ applied at time $t=0$ ($\sigma=45$, 50, 52, 53, 55, 60, 65, 70, and 75~Pa from right to left, adapted from \cite{Grenard:2014} with permission from the Royal Society of Chemistry). (b)~Fatigue experiments in a carbon black gel at 6~wt~\% in light mineral oil: strain amplitude response $\gamma_0(t)$ to oscillatory stresses of frequency $f=1$~Hz and amplitude $\sigma_0=11$, 15 and 27~Pa from right to left. Inset: $G'$ (solid line) and $G''$ (dashed line) as a function of time $t$ for $\sigma=11$~Pa (adapted from \cite{Perge:2014b} with permission from the Society of Rheology and \cite{Gibaud:2016} with permission from the Royal Society of Chemistry). (c)~Fluidization time $\tau_f$ as a function of stress $\sigma$ in creep experiments ($\bullet$) and as a function of stress amplitude $\sigma_0$ in fatigue experiments ($\blacksquare$) for a carbon black gel at 6~wt~\% in light mineral oil. Lines corresponds to the best exponential fits of the data: $\tau_f\sim \exp(-\sigma/\sigma^*)$ with $\sigma^*=7$~Pa for fatigue and $\sigma^*=3.5$~Pa for creep experiments (adapted from \cite{Grenard:2014} and \cite{Gibaud:2016} with permission from the Royal Society of Chemistry). {\bf (d)--(f)~Brittle fracture in sodium caseinate gels}. (d) Creep experiments in an aqueous gel of sodium caseinate at 4~wt~\% acidified with 1~wt~\% GDL: shear rate responses $\dot\gamma(t)$ for different shear stresses $\sigma$ applied at time $t=0$ ($\sigma=200$, 300, 400, 550, and 1000 Pa from right to left, adapted from \cite{Leocmach:2014} with permission from the American Physical Society). (e)~Fatigue experiments in an aqueous gel of sodium caseinate at 6~wt~\% acidified with 6~wt~\% GDL : strain amplitude response $\gamma_0(t)$ to oscillatory stresses of frequency $f=1$~Hz and amplitude $\sigma_0= 90$, 150 and 200~Pa from right to left (adapted from \cite{StMichel:2017} with permission from the Royal Society of Chemistry).
         (f)~Fluidization time $\tau_f$ as a function of stress $\sigma$ in creep experiments ($\bullet$) and as a function of stress amplitude $\sigma_0$ in fatigue experiments ($\blacksquare$) for a sodium caseinate gel at 4~wt~\% in water acidified with 1~wt~\% GDL. Lines correspond to the best power-law fits of the data: $\tau_f\sim \sigma^{-\beta}$ with $\beta=4.2$ for fatigue and $\beta=5.5$ for creep experiments (unpublished data).}
     \label{fig:cascbcreep} 
 \end{figure}

The yielding dynamics of carbon black gels associated with creep tests and fatigue tests is presented in Fig.~\ref{fig:cascbcreep}(a)--(c). In both cases, the yielding transition is strongly time- and load-dependent. Figure~\ref{fig:cascbcreep}(a) shows the evolution of the shear rate $\dot{\gamma}$ as function of time $t$ under constant stress amplitudes $\sigma$ (increasing from right to left). All measurements show similar features: $\dot{\gamma}$ first slowly decreases at short times, then shows an abrupt increase and presents one or several kinks and fluctuations before a final increase up to a steady state. The whole process takes from a few seconds for large stresses to several hours at low stresses. The fluidization time $\tau_f$ is defined as the inflection point during the final increase of $\dot{\gamma}$. Measurements of the velocity field within the gel revealed that the inflection point indeed corresponds to the point where the sample flows homogeneously like a liquid \cite{Gibaud:2010,Grenard:2014}. Similarly, Fig.~\ref{fig:cascbcreep}(b) displays the temporal evolution of the strain amplitude $\gamma_0$ in response to three different fatigue tests (i.e., oscillatory shear stress at constant stress amplitude) conducted at various amplitudes $\sigma_0$. Here again, all measurements show the same trend: a slow initial increase of $\gamma_0$ followed by a final increase up to a steady state. As shown in the inset of Fig.~\ref{fig:cascbcreep}(b), the final increase of $\gamma_0$ corresponds to the transition from a solid-like state where $G'>G''$ to a fluid-like state where $G''>G'$. The fluidization time $\tau_f$ is here defined as the inflection point of the final increase of $\gamma_0$, which corresponds to the time where the two moduli roughly coincide, $G'\simeq G''$, at least for low enough stress amplitudes [see inset in Fig.~\ref{fig:cascbcreep}(b)]. The time-dependence observed here is usually referred to as ``delayed flow'' or ``delayed yielding'' in the literature \cite{Gopalakrishnan:2007,Sprakel:2011,Lindstrom:2012}. Finally, note that in both creep and fatigue tests, a steady state is reached where the carbon black dispersion flows like a liquid. When the test is stopped and the stress is released, the dispersion aggregates again into a gel. As a result, successive yielding experiments can be performed on a single sample provided that the same preshear protocol is repeated between each test to prepare the sample into the same initial state.  

Based on series of creep and fatigue tests, one may then access the dependence of $\tau_f$ with the applied stress. As shown in Fig.~\ref{fig:cascbcreep}(c), such a dependence is well captured by an exponential decrease, $\tau_f=\tau_0 \exp(-\sigma/\sigma^\star)$, for both types of tests, although with different characteristic stresses $\sigma^\star$ and prefactors $\tau_0$. Such an empiric law was first reported for carbon black gels in \cite{Gibaud:2010,Sprakel:2011}. It is well captured by a mean-field model based on Kramers’ theory for activated processes \cite{Sprakel:2011,Lindstrom:2012,Gibaud:2016}. In this framework, yielding results from the competition between individual interparticle bond formation and rupture events induced by the external stress. Moreover, the characteristic stress $\sigma^\star$ reflects the mean elastic barrier that should be overcome for a single bond to break. Typical values of $\sigma^\star$ in carbon black gels range from 1 to 10 Pa. We note from Fig.~\ref{fig:cascbcreep}(c) that this stress barrier is smaller in creep experiments than in fatigue experiments ($\sigma^*\simeq3.5$~Pa vs. $\sigma^*\simeq 7$~Pa resp.). Such an Arrhenius-like behavior of $\tau_f$ is observed in many other types of colloidal gels that also show reversible yielding, including depletion gels \cite{Ramakrishnan:2005,Dibble:2006,Koumakis:2015,Sprakel:2011}, presheared proteins gels \cite{Brenner:2013}, and thermo-reversible gels \cite{Gopalakrishnan:2007}. As for carbon black gels, they appear as a model system to investigate delayed yielding. For the sake of simplicity, we only focus here on $\tau_f$. Yet, we emphasize that the fluidization process also generically involves several other time scales, e.g., due to the early detachment from the shearing walls at large stresses, and that $\tau_f$ also depends on boundary conditions as well as on the preshear protocol or on external mechanical vibrations. For full details, the reader is referred to \cite{Gibaud:2010,Grenard:2014,Perge:2014b,Gibaud:2016,Helal:2016,Gibaud:2020}.

\subsection{Caseinate gels as model materials for irreversible yielding}
\label{sec:casein_fracture}

The counterpart of Fig.~\ref{fig:cascbcreep}(a)--(c) for sodium caseinate gels is displayed in Fig.~\ref{fig:cascbcreep}(d)--(f). Here again, the strong time- and load-dependence of the yielding process is evident. In the creep experiments of Fig.~\ref{fig:cascbcreep}(d), the shear rate first decreases as a power law of time $\dot{\gamma} \sim t^{-\alpha}$, with $\alpha \simeq 0.85$. This initial regime is strongly reminiscent of the primary creep regime observed in hard solids \cite{Miguel:2002,Nechad:2005,Rosti:2010} and referred to as ``Andrade creep'' \cite{Andrade:1910}. Here, the value of $\alpha$ observed in sodium caseinate gel can be inferred from linear viscoelasticity \cite{Leocmach:2014}, and quantitatively differs from the value $2/3$ classically reported in hard solids, where such a primary creep response is interpreted in terms of collective plastic events \cite{Miguel:2008}. Then $\dot{\gamma}$ reaches a minimum and finally diverges at the fluidization time $\tau_f$. In the fatigue experiments shown in Fig.~\ref{fig:cascbcreep}(e), the strain increases slowly before diverging upon yielding, which defines the fluidization time $\tau_f$. The striking divergence of $\dot\gamma(t)$ and $\gamma_0(t)$ is the hallmark of irreversible rupture in the investigated sodium caseinate gels. Contrary to the case of reversible yielding where a saturation is observed, the velocity of the moving tool keeps increasing for all applied stresses until the rheometer reaches its speed limit. In the process, the gel structure is fully destroyed. Moreover, the gel does not recover once the load is released, which is why $\tau_f$ corresponds to the ``fracture time'' \cite{vanVliet:1995}. As a consequence, one has to perform the various yielding tests on different samples, each being freshly prepared for a given set of parameters. 

Furthermore, as displayed in Fig.~\ref{fig:cascbcreep}(f), the rupture time does not follow an exponential scaling but its dependence with stress is rather given by a power law,  $\tau_f\sim\sigma^{-\beta}$, with $\beta=5.5$ under creep and $\beta=4.2$ under fatigue. This power-law scaling was first reported for sodium caseinate gels by \cite{Leocmach:2014} and is strikingly reminiscent of the Basquin law of fatigue \cite{Basquin:1910} found for a variety of heterogeneous or cellular materials under cyclic deformation \cite{Kohout:2000,Kun:2008}. This behavior can be recovered from a fiber bundle model where the system under stress is represented by a set of linear elastic fibers following simple failure rules \cite{Jagla:2011}. When the fiber bundle is subjected to an increasing external load, the fibers behave like linear elastic springs until they break for a given failure load. To recover the Basquin law, \cite{Kun:2008} modified the original fiber bundle model and considered a mean-field approach in which the network strands form a random network and fail either due to immediate breaking or to aging through damage accumulation. In this framework, $\beta$ is directly related to the growth of local damage as a function of the local stress. High values of $\beta$ mean that the material is prompt to accumulate damage. In soft gels, typical values of $\beta$ range from 1 to 10, much larger that in hard solids such as metals ($\beta\simeq0.1$) \cite{eshbach1990} and asphalt ($\beta\simeq0.5$) \cite{Kun:2008}. Here, the larger exponent found in creep tests than in fatigue experiments suggests that sodium caseinate gels accumulate damage more ``efficiently'' under a constant stress than under an oscillatory stress. In any case, both the strain divergence during the approach of yielding and the Basquin power-law scaling for the rupture time are indicative of a brittle-like behavior of the strand network, which strongly differs from the delayed yielding observed in gels constituted of colloidal clusters. This difference is further explored through more local approaches in the next paragraph.

\subsection{Characterizing yielding through spatially-resolved dynamical measurements}
\label{sec:local}

Measuring the functional form $\tau_f(\sigma)$ constitutes a big step forward when compared to the simple amplitude-sweep measurements or to the mere estimation of a yield stress or strain. In particular, the stress-dependence of the fluidization time allows one to classify the yielding behavior, to infer a rupture mechanism and to quantitatively test models, e.g., based on reversible bond-breaking or on damage accumulation. This approach remains however macroscopic and more local measurements are required to better pinpoint the rupture mechanism. Such measurements can be performed at different length scales: at the mesoscopic scale, typically 10--100~$\mu$m, in order to map the displacement field within the gel or at the microscopic scale, typically 0.1--10~$\mu$m, to directly retrieve the structure of the strands or clusters during fatigue or creep experiments.

 \begin{figure}
 	\centering
 	\includegraphics[width=1\columnwidth]{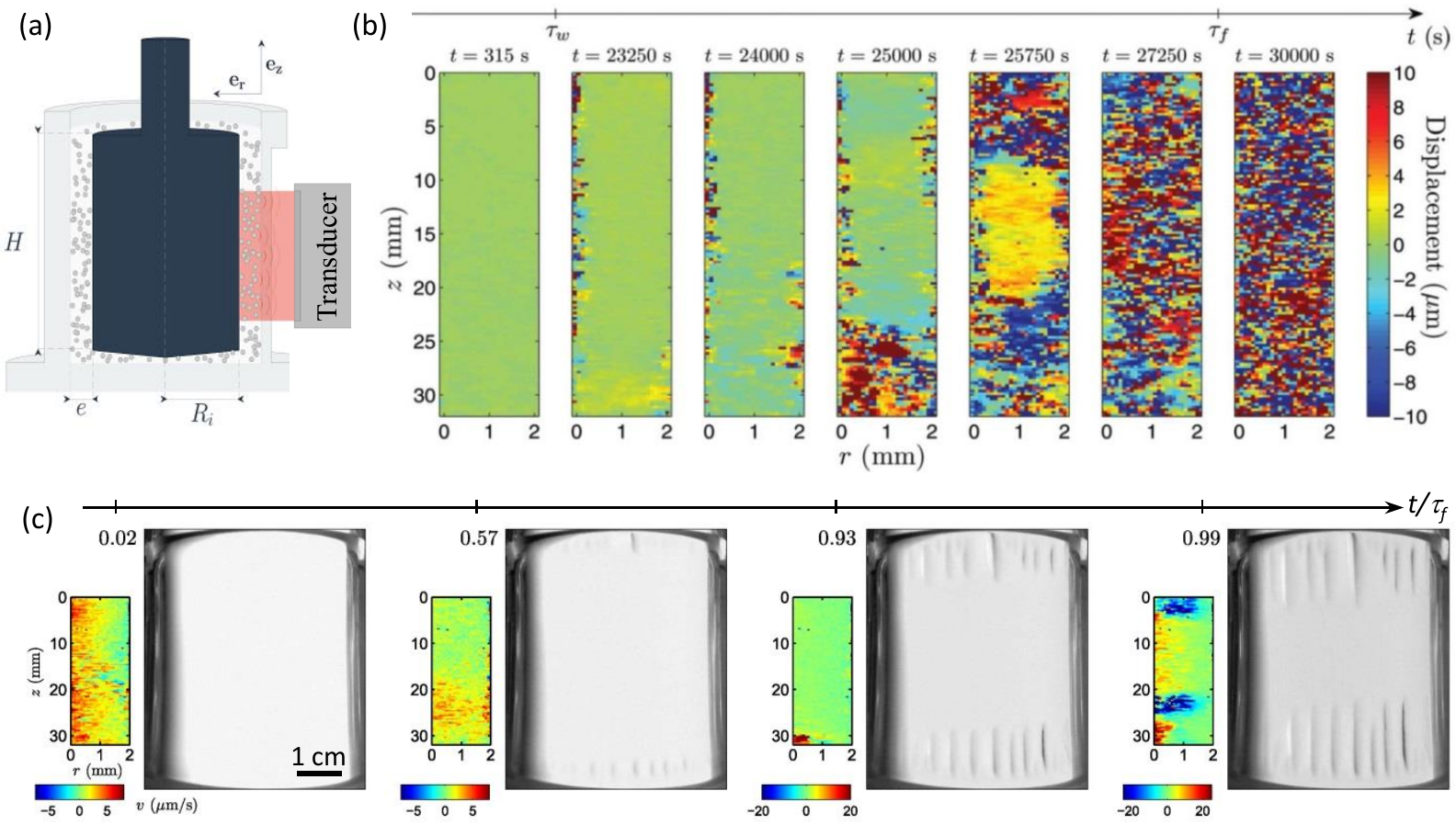}
         \caption{{\bf Heterogeneous dynamics during yielding of carbon black and sodium caseinate gels.} (a) Sketch of the rheology experiment coupled to high-frequency ultrasonic imaging. The rheology is carried out using a Taylor-Couette geometry, a concentric-cylinder geometry of height $H$ and of gap $e$  with an inner rotating cylinder of radius $R_i$.  (b)~Time-resolved displacement maps $\Delta(r,z,t)$ recorded in a carbon black gel at 6~wt~\% in light mineral oil between two cycles of oscillation during a fatigue experiment with $\sigma_0=11$~Pa and $f=1$~Hz (adapted from \cite{Perge:2014b} with permission from the Society of Rheology). The experiment is performed in the experimental setup sketched in (a) with $H=50$ mm, $e=2$ mm and $R_i=48$ mm. $r$ denotes the radial distance from the inner cylinder and $z$ the position along the height of the cylinder. $\tau_w$ indicates the time at which the gel detaches from the inner cylinder and $\tau_f$ indicates full bulk fluidization, which corresponds to the last inflection point of $\gamma_0(t)$ in Fig.~\ref{fig:cascbcreep}(b). (c)~Typical fracture growth in a sodium caseinate gel at 4~wt~\% acidified with 1~wt~\% GDL and subjected to a constant shear stress $\sigma=300$~Pa in the experimental setup sketched in (a) with $H=60$ mm, $e=2$ mm and $R_i=23$ mm. The picture taken at $t/\tau_f=0.57$ corresponds to the time at which the shear rate passes through a minimum. Velocity maps $v(r,z,t)$ recorded simultaneously to both the rheological data and to the images are shown next to the corresponding picture (adapted from \cite{Leocmach:2014} with permission from the American Physical Society). $\tau_f$ is the ``fracture time'' defined from rheology as the point where $\dot\gamma(t)$ diverges in Fig.~\ref{fig:cascbcreep}(d) and which also corresponds to when fractures meet in the middle of the cell.}
     \label{fig:cbspatiotemporel}
 \end{figure}

At the mesoscopic scale, displacement maps and velocity profiles can be obtained from scattering-based techniques \cite{Adam:1988,Kroon:1996,Romer:2000,Duri:2005}, magnetic resonance imaging \cite{Assink:1991,Bonn:2008}, image correlation microscopy \cite{Derks:2004,Larsen:2008,Edera:2017} or ultrasonic imaging \cite{Manneville:2004a,Gallot:2013,StMichel:2016}. When coupled to rheology, the latter high-frequency ultrasonic imaging technique, sketched in Fig.~\ref{fig:cbspatiotemporel}(a), allows one to measure the displacement map within the gel with a spatial resolution of about $50~\mu$m simultaneously to standard rheological observables. Compared to light-based techniques, ultrasound has the great advantage of being able to probe optically opaque gels, in spite of lower spatial resolution. In the case of the carbon black and sodium caseinate gels investigated so far, which are transparent to ultrasound, this imaging technique necessitates to seed the gel with acoustic contrast agents, respectively hollow glass spheres of mean diameter 6~$\mu$m and polyamide spheres of mean diameter 30~$\mu$m. 

Figure~\ref{fig:cbspatiotemporel}(b) shows displacement maps recorded during a fatigue experiment carried out on a carbon black gel. One can identify four successive regimes. (i)~At short time scales, the gel displacement from one oscillation cycle to the other is zero, i.e., the gel behaves like an elastic solid. (ii)~For $t=\tau_w$, the gel yields at the inner cylinder of the geometry (located at $r=0$) inducing a plug-like flow of the bulk gel. (iii)~The gel gets progressively fragmented into large solid domains (that show as uniform patches on the displacement map) evolving in a fluid background (that shows as apparently noisy regions where the displacements are very large and uncorrelated from one pixel to the other). (iv)~Finally, at $t=\tau_f$, those large domains get fully eroded, and the whole gel is turned into a homogeneous fluid. As described in detail in \cite{Perge:2014b,Gibaud:2016}, this last step corresponds to the final increase in the strain amplitude $\gamma_0$ reported in Fig.~\ref{fig:cascbcreep}(b).

Thus, despite a complex yielding process characterized by solid-liquid coexistence and heterogeneous flows, the  carbon black gel ends up being homogeneously fluidized. The case of a sodium caseinate gel is strikingly different. Figure~\ref{fig:cbspatiotemporel}(c) shows pictures of the gel subjected to a creep experiment in a concentric-cylinder cell together with time-resolved velocity maps. Three regimes are observed under a constant stress. (i)~During primary creep, the gel is linearly deformed and behaves like an elastic solid. (ii)~At intermediate times, around the time at which $\dot\gamma(t)$ reaches a minimum value in Fig.~\ref{fig:cascbcreep}(d), regularly-spaced cracks start to nucleate from the top and bottom edges of the cell. (iii)~Finally, cracks grow along the vorticity direction, i.e., perpendicular to the applied stress, and the gel eventually fails when fractures meet in the middle of the cell at $\tau_f$, which translates into the divergence of the shear rate. The macroscopic fractures within the sodium caseinate gel network are invaded with water expelled from the gel matrix. Therefore, the yielding behavior of the present acid-induced protein gel can be described as a stress-induced macroscopic fracture of the gel network leading to an irreversible separation of the sample between a dilute protein suspension and broken pieces of the original gel.

To summarize, we have illustrated how reversible and irreversible yielding processes differ at both macroscopic and mesoscopic scales based on optical and ultrasound imaging. Note that the gel microstructure under stress can also be probed at scales below 10~$\mu$m by coupling the shearing device to scattering techniques such as small-angle neutron, x-ray or light scattering \cite{Verduin:1996,Pignon:1997a,Varadan:2001,Mohraz:2005,Vermant:2005,Eberle:2012,Kim:2014,Hipp:2019}. When time-resolved measurements are available, such techniques provide valuable insight into the yielding dynamics in the reciprocal space, i.e., as a function of the scattering wave number. For instance, they allow one to identify the relevant length and time scales of plastic activity \cite{AIme:2018}. However, they fail to provide a direct picture of the microscopic mechanisms underpinning the physics of shear-induced yielding in colloidal gels. As discussed below in Section~\ref{sec:micro}, fast confocal microscopy appears as an even more promising alternative to tackle such a challenge.

\section{Fundamental questions and perspectives}
\label{sec:discussion}

In the previous sections, we have illustrated how one may investigate the yielding transition on two specific cases representative of two distinctive behaviors, namely reversible fluidization and irreversible fracture. Such a distinction opens a number of fundamental questions, e.g., on the predictability of yielding or on the microscopic rupture events pre-existing macroscopic failure. In the following section, we list such questions and draw a few perspectives for future work.

\subsection{What is the physical meaning of the functional form $\tau_f(\sigma)$?}

We have identified two types of functional forms for the empirical failure law $\tau_f(\sigma)$, namely an exponential in the case of carbon black gels, and a power law in the case of sodium caseinate gels. The reason for such a striking difference and the physical origins of these laws remain unclear and raise several outstanding questions. First, does such a difference at the macroscale directly stem from the fact that bonds can spontaneously reform in carbon black gels, whereas they cannot in sodium caseinate gels, or does it originate only from stress-induced plastic events at the microscale? In that framework, one may ask whether precursors to failure are of different nature for these two functional forms, and thus if they could help anticipate the nature of the failure scenario thanks to a single creep or fatigue experiment. Second, one may be tempted to rationalize these two functional forms as two cases of a single master function $\tau_f(\sigma)$ that could involve a Weibull distribution for instance~\cite{vanderzwaag:1989}. This raises the question of whether it is possible to go continuously from one type of functional form to the other within the same type of colloidal gel and how to devise such an experimental system.
 
 \subsection{How does strain-induced yielding compare to stress-induced yielding?}
 
So far, we have focused on stress-induced yielding through both creep and fatigue experiments. Another way to induce yielding is by imposing a constant \textit{shear rate}, $\dot\gamma$, in a shear start-up experiment as sketched in Fig.~\ref{fig:test}(d). Such a flow test amounts to submitting the material to a shear strain $\gamma(t)=\dot\gamma t$ that increases linearly with time. As long as the deformation lies in the linear domain, imposing the strain is equivalent to imposing the stress. However, such an equivalence \textit{a priori} breaks down when the material enters the nonlinear regime so that a natural question is how yielding under a given shear rate may be compared to that induced by stress and whether it is possible to relate the two.

\begin{figure}[tbh]
 	\centering
 	\includegraphics[width=1\columnwidth]{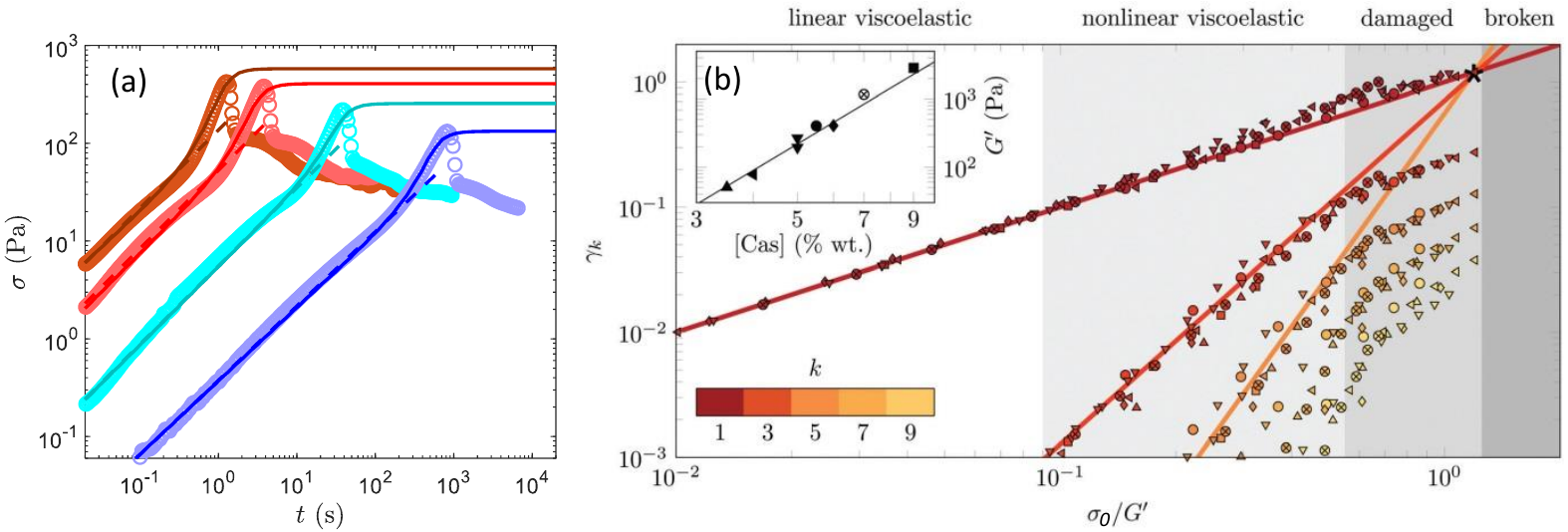}
         \caption{(a)~{\bf Strain-induced rupture.} Shear stress response $\sigma$ vs time $t$ of a sodium caseinate gel at 4~wt~\% acidified with 1~wt~\% GDL and subjected to a constant shear rate $\dot\gamma = 10^{-3}$, 0.02, 0.2 and 0.6 s$^{-1}$ (from right to left) applied at $t = 0$. The fracture time $T_f$ corresponds to the locus of the stress maximum. Dashed lines correspond to linear viscoelastic response and solid lines show the K-BKZ prediction (adapted from \cite{Keshavarz:2017} with permission of the American Chemical Society). (b)~{\bf Rupture prediction from Fourier transform rheology.} Amplitude of the Fourier modes $\gamma_k$ of the strain response of sodium caseinate gels subjected to an oscillatory stress of amplitude $\sigma_1$ for various casein concentrations. Colors code for $k = 1$ to 9. The stress amplitude is normalized by the elastic modulus $G'$ of the gel measured in the linear viscoelastic regime. Solid lines show the best power-law fits of $\gamma_k$  for $\sigma_1/G'<0.5$. The build-up of Fourier modes within the nonlinear viscoelastic regime yields an empirical criterion for rupture as the intersection point of the power-law extrapolations into the damaged regime. Inset: elastic modulus $G'$ as a function of the sodium caseinate weight concentration [Cas]. Symbols code for casein concentrations in the main graph. The solid line is $G' \sim [{\rm Cas}]^4$ (adapted from \cite{StMichel:2017} with permission from the Royal Society of Chemistry).}
     \label{fig:prediction}
 \end{figure}

Figure~\ref{fig:prediction}(a) shows the stress responses of a sodium caseinate gel during shear start-up experiments performed under various shear rates. The general shape of $\sigma$ vs $t$ resembles that of $\sigma_0$ vs $\gamma_0$ reported in Fig.~\ref{fig:rheoyield}(b). In particular, it presents a marked stress increase characteristic of strain stiffening prior to the sudden drop that marks the sample macroscopic failure. As explained at length in \cite{Keshavarz:2017}, such stress responses can be quantitatively modelled by a K-BKZ (Kaye–Bernstein–Kearsley–Zapas) constitutive formulation \cite{Bird:1987,Larson:1999} based on the linear viscoelastic spectrum and a ``damping function'' inferred empirically from step-strain experiments. The predictions of this K-BKZ approach are displayed as solid lines in Fig.~\ref{fig:prediction}(a) and closely match the experimental data up to the rupture point. Interestingly, one may relate the fracture time $T_f(\dot\gamma)$, which is now a function of $\dot\gamma$ since measured under a constant shear, to the Basquin law $\tau_f(\sigma)$ obtained from creep tests thanks to the Bailey ``durability'' criterion first introduced for glasses and metals \cite{Bailey:1939,Freed:2002} and later applied to the rupture of polymeric liquids and elastomers \cite{Eirich:1972,Malkin:1997}. This criterion states that $\int_0^{T_f}{\rm{d}t}/{\tau_f[\sigma(t)]}=1$, where the creep rupture time $\tau_f$ is evaluated all along the loading process $\sigma(t)$. The Bailey criterion assumes that the failure of the material results from the accumulation of independent local failure events. Combining the K-BKZ approach with the Bailey criterion allows one to quantitatively predict the scaling of the yield stress and yield strain as a function of the applied shear rate $\dot{\gamma}$. This result nicely extends the range of application of Bailey's criterion to soft viscoelastic gels, further reinforcing the emerging analogies between soft and hard materials \cite{Schmoller:2013}.
 
Still, the above approach to reconcile stress-induced and strain-induced failure requires a careful modeling of the linear and nonlinear viscoelastic response. In the specific case of sodium caseinate gels, it is successful thanks to a robust, well-defined kernel for the K-BKZ equation. Besides, in these gels, the deformation remains homogeneous well into the nonlinear viscoelastic regime without significant structural damage. In practice, however, the gel may detach from the wall and/or present heterogeneous deformation fields, as already shown above in Fig.~\ref{fig:cbspatiotemporel}(b) in the case of carbon black gels. Therefore, it is not clear whether a similar approach, possibly based on a different durability criterion, is achievable in the case of reversible, delayed yielding due to activated processes, or even in other colloidal gels showing irreversible, brittle-like rupture. Finally, the Bailey criterion necessitates previous knowledge of the full $\tau_f(\sigma)$ law and, as such, does not allow one to become genuinely predictive of rupture. This observation leads us to ask whether it is at all possible to predict yielding in colloidal gels based on the sole measurement of macroscopic rheological observables.

\subsection{Can one predict the yielding scenario from macroscopic observables?}

Rupture prediction consists in using indicators of the material properties way before it fails to predict the time, strain, or stress at which the sample will fail. Establishing a yielding prediction from the sole rheological measurements appears as a major challenge for colloidal gels. \cite{Leocmach:2014} have shown that the time $\tau_{\rm min}$ at which the shear rate reaches its minimum in the creep experiments of Fig.~\ref{fig:cascbcreep}(d) is the same fraction of the rupture time $\tau_f$ whatever the applied stress, namely $\tau_{\rm min}\simeq 0.57\tau_f$. Such a proportionality is known as the Monkmann-Grant relation \cite{Monkman:1956,Voight:1989} and has been reported in a broad variety of  solids \cite{Sundararajan:1989,Nechad:2005,Laurson:2011}. It allows for a rough estimate of the rupture time by recording $\dot\gamma(t)$ up to the point where it reaches a minimum. Still, sample to sample variations usually dominate, which limits the predictions of the failure time to within at least 10\% \cite{Koivisto:2016}. Moreover, as seen in Fig.~\ref{fig:cbspatiotemporel}(c), at this point, macroscopic fractures have appeared, and the network structure is already deeply affected by stress.

For the same sodium caseinate gels, Fig.~\ref{fig:prediction}(b) shows that, based on Fourier transform rheology \cite{Wilhelm:2002}, a prediction of rupture under an oscillatory stress can be empirically achieved before any significant damage occurs. There, we take advantage of the increasingly rich harmonic content of the strain response to an oscillatory stress of increasing amplitude \cite{StMichel:2017}. The strain response is decomposed into a Fourier series, $\gamma(t) = \sum_k \gamma_k \cos \left ( 2 \pi k f t + \phi_k \right ) $, where $\gamma_k$ is the amplitude of the $k^{\rm th}$ harmonics and $\phi_k$ its phase with respect to the sinusoidal stress input $\sigma(t)=\sigma_1\cos( 2 \pi f t)$, which is applied for only 10 cycles of oscillations and for increasing amplitude $\sigma_1$. Between two sets of oscillation cycles, the gel is left to recover for 10~min. The amplitudes of the harmonics $\gamma_{k>1}$ follow power laws of the applied stress amplitude with exponents that increase with $k$ and that are larger than 1 in the non-linear viscoelastic regime and eventually get close to 1 in the damaged regime. The extrapolations of the power laws from the non-linear viscoelastic regime all intersect at a single point. Interestingly, this single point matches remarkably well the experimental rupture point ($\sigma_c/G' = 1.06 \pm 0.25, \gamma_c= 1.32 \pm 0.20$). This empirical analysis seems to point out that rupture is encoded in the non-linear strain modes and that, for sodium caseinate gels, a simple extrapolation of Fourier modes allows one to \textit{predict} the rupture point way before the gels actually fail. Such an idea echoes findings from Fourier transform rheology on other systems such as polymer melts \cite{hirschberg:2017}.

Here again, the fact that the deformation field remains homogeneous and that the gel sticks to the walls deep into the nonlinear regime might help to get a simple rupture prediction in the present protein gel. Furthermore, in the absence of any theoretical basis, such a prediction appears as an ad-hoc criterion, and more work is needed to understand its origin. Whether or not a similar predictive approach, either based on Fourier modes or on more advanced indicators from LAOS measurements \cite{Ewoldt:2008,Dimitriou:2013}, can be proposed for reversible yielding of colloidal gels stands out as an open question.

\subsection{What does strain-stiffening tell us about yielding?}
\label{sec:stiffening}
 
\begin{figure}[t!]
 	\centering
 	\includegraphics[width=1\columnwidth]{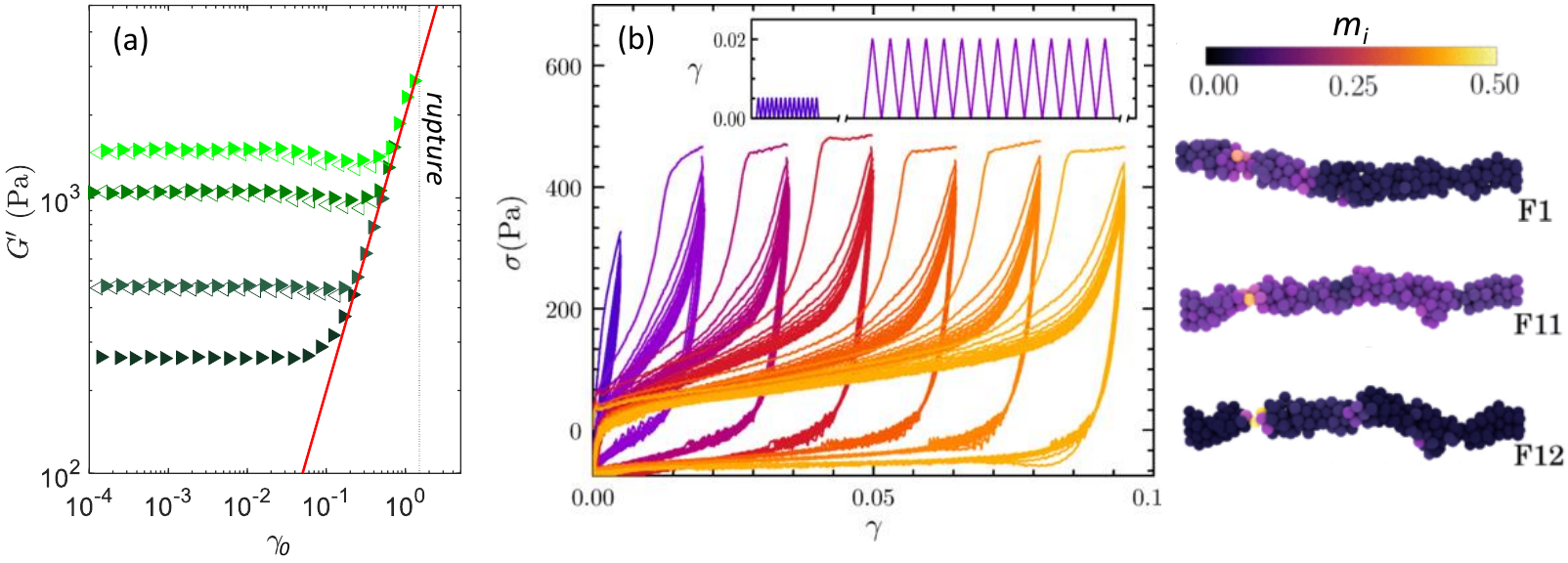}
         \caption{{\bf Strain stiffening prior to rupture.} (a)~Response of a natural rubber latex colloidal gel at a volume fraction of 0.11 to successive amplitude sweeps under oscillatory shear. Each cycle consists of a continuous increase in strain amplitude $\gamma_0$ (full symbols) up to a predetermined maximum strain $\gamma_{\rm max}$, followed  by a continuous decrease in strain amplitude (empty symbols). Each new cycle is performed on the same sample. The elastic modulus increases irreversibly until rupture at $\gamma_c\simeq 1.3$. The red line highlights the strain stiffening behavior the gel, $G'\sim \gamma_0^{1.0}$ (adapted from \cite{dereis:2019} with permission from the Royal Society of Chemistry). (b)~Stress-strain curves $\sigma(t)$ vs $\gamma(t)$ of an attractive polystyrene gel at a volume fraction of 0.18 for a triangular strain input with $\dot \gamma =0.1$~s$^{-1}$ and increasing values of the maximum strain $\gamma_{\rm max}=0.005$, 0.02, 0.035, 0.05, 0.065, 0.08, 0.095 (from blue to yellow, see also inset). The right panel shows visual representations of the plastic deformation in a computer-simulated gel strand after the first cycle (F1), before fracture (F11), and after fracture (F12). Colors code for low (purple) to high (yellow) noncumulative plastic deformation $m_i$ per particle (adapted from \cite{vandoorn:2018} with permission from the American Physical Society).}
     \label{fig:strainstiff}
 \end{figure}

Related to the previous question of the yielding predictability is the observation that colloidal gels do not all become weaker before rupture. Although strain softening corresponds to the most general case, as exemplified in Fig.~\ref{fig:rheoyield}(a) for carbon black gels, some gels show the reverse trend. Indeed, their elastic modulus $G'$ increases as the strain or stress increases, i.e., these gels stiffen until abrupt rupture takes place. Such a strain-stiffening behavior has also been reported in a wide range of materials including rubbers, swollen polymer gels and semi-flexible biopolymer networks such as actin~\cite{Gardel:2004,Schmoller:2010}, heat-set $\beta$-lactoglobulin~\cite{Pouzot:2006}, collagen \cite{Arevalo:2015}, fibrin, and vimentin or neurofilaments~\cite{storm:2005}. In colloidal gels, strain stiffening is rather rare and, when present, the amplitude of the phenomenon is usually small with $G'$ increasing at most by a factor of 5. This is the case for polystyrene latex particles that gel under the addition of MgCl$_2$~\cite{gisler:1999} or for the sodium caseinate gels that we have been discussing throughout in this article, as displayed in Fig.~\ref{fig:rheoyield}(b). Recently, however, it was reported that natural rubber latex gels~\cite{dereis:2015} and thermosensitive polystyrene colloidal gels~\cite{vandoorn:2018} display a strain-stiffening behavior with an amplitude comparable to that of biopolymer gels for which $G'$ may increase by a factor of 100. As shown in Fig.~\ref{fig:strainstiff}(a) and (b), the elasticity of these gels adapts to the imposed stress or strain and grows up to rupture.

The interpretation of such an impressive strain stiffening and its relationship with subsequent yielding are not yet settled. Using numerical simulations, \cite{vandoorn:2018} suggest that plasticity at the strand level is responsible for the strain-stiffening behavior. Mechanical loading leads to strand stretching, which displays a strong hysteresis: it builds slack into the network that softens the solid at small strains but causes strain stiffening at larger deformations. \cite{bouzid:2017} have recently tackled this issue through computer simulations and found that strain stiffening could be understood in terms of backbone network topology: the stiffening is observed in stiffer gels where there is a lack of soft bending relaxation modes. Those gels exhibit purely stiffening behavior due to their still relatively sparse backbone connectivity that allows for stretching and orienting the gel branches along the direction of maximum elongation in shear.

Still, natural rubber latex gels must fall in another category as the strain-stiffening process is irreversible \cite{dereis:2019}: once stiffened the material never regains its initial elasticity. Here, it can be hypothesized that the network is not just stretched and oriented but that it restructures, most probably through strand coalescence along the shear direction. Another interpretation could rely on the details and specificity of the particles themselves. In any case, it is clear that the strain-stiffening phenomenon that precedes rupture deserves renewed attention as it may help discriminate between various plastic responses or restructuring processes in some colloidal gels. 

\subsection{Can we get a microscopic picture of yielding?}
\label{sec:micro}

 \begin{figure}[b!]
 	\centering
 	\includegraphics[width=1\columnwidth]{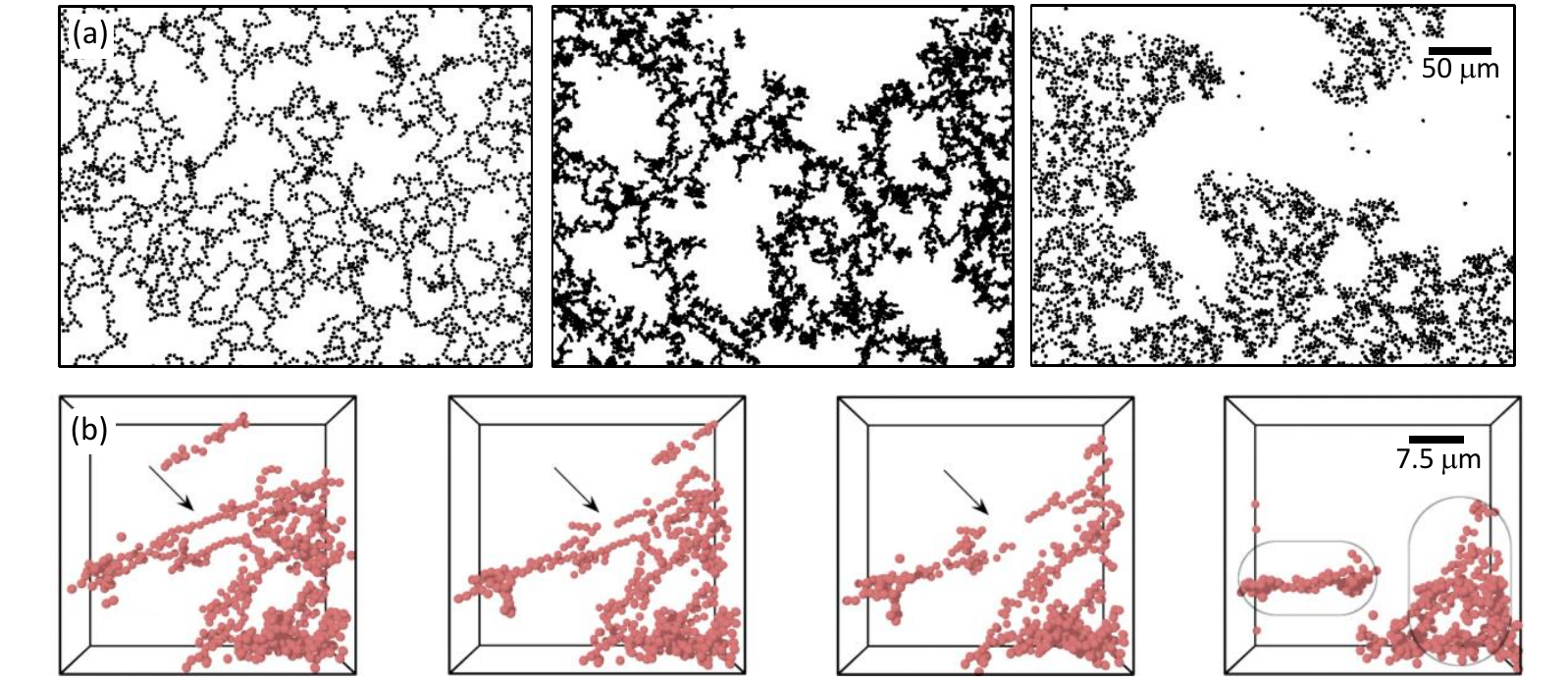}
         \caption{{\bf Shear-rate-induced rupture at the microscopic level.} (a)~Microscopy images of a two-dimensional colloidal gel composed of 3.1 $\mu$m polystyrene particles adsorbed at an oil–water interface with a surface fraction of 0.3. From left to right: gel structure observed just after aggregation  and during steady-state flows at $\dot{\gamma}= 0.015$ and 0.026~s$^{-1}$. Note that particles have been reduced by about 20\% in size for better visualisation (adapted from \cite{Masschaele:2011} with permission from the Royal Society of Chemistry). (b)~Three-dimensional images reconstructed from a confocal microscopy experiment of a gel composed of 1~$\mu$m sterically-stabilized PMMA colloids at a volume fraction of 0.035 in an index-matched solvent. Polymers where added to the dispersion to mediate an attraction between the colloids (depletion interactions) and form a gel. From left to right: time sequence of a local rupture event indicated by the arrow, which leads to the formation of two distinct clusters highlighted by circles. The imposed shear rate is $\dot{\gamma}= 0.07$~s$^{-1}$. The total duration of the sequence is about 100~s (adapted from \cite{Rajaram:2010} with permission from the Royal Society of Chemistry).}
     \label{fig:confocal}
 \end{figure}
 
As already pointed out above in Section~\ref{sec:local}, macroscopic observables such as rheological data, although very informative, do not tell us anything about the microscopic events that govern yielding in colloidal gels. Therefore, additional experimental tools are needed to explore yielding at the cluster or strand scales. To be able to anticipate failure, one needs to identify \textit{precursors} of yielding both for reversible and irreversible yielding. Such precursors are broadly defined by \cite{Cipelletti:2020} as ``any sharp variation of a parameter quantifying the system structure or dynamics on length scales of the order of, or slightly larger than, the relevant length scale of the sample structure, [and that] should be detectable well in advance of macroscopic failure.'' A major step forward has been achieved recently by \cite{AIme:2018} in the case of a gel of silica colloids that shows irreversible rupture. By coupling rheometry to small-angle light scattering, it was shown that the minimum in the shear rate corresponds to a sudden burst of plastic activity that is interpreted as the proliferation of irreversible rearrangements most probably consecutive to bond-breaking events. This dynamic precursor irreversibly weakens the network until it fully breaks.

Although very promising, the previous dynamic light-scattering technique still fails to provide detailed, spatially-resolved information on the individual events that are responsible for plastic activity. To access such information, direct microscopic visualization must be performed. To date, only a handful of studies have addressed this question in colloidal gels. In particular, monolayers of attractive colloidal particles trapped at an liquid--gas interface or at a liquid--liquid interface have allowed one to image model two-dimensional gels with video microscopy under strain or stress and to study flow-induced anisotropy in the aggregated layer \cite{Hoekstra:2003,Vermant:2005}. It was shown by \cite{Masschaele:2009,Masschaele:2011} that yielding is preceded by a cascade of bond-breaking events, which number steadily increases up to a strain $\gamma\sim 1$. Dangling chains that are formed upon breaking may either fold back to the backbone or reconnect with other dangling ends. In the former case, the backbone densifies while connectivity is restored in the latter case.  For $\gamma> 1$, global connectivity is lost, and one is left with a more heterogeneous structure than that of the gel at rest, with larger voids that coexist with more compact clusters as illustrated in Fig.~\ref{fig:confocal}(a).

In three-dimensional gels, confocal laser scanning microscopy (CSLM) as emerged as a powerful tool to study colloidal gels under flow \cite{Prasad:2007}. Pioneering experiments by \cite{Varadan:2003b} on thermoreversible gels of silica colloids showed that squeeze flow induces voids at moderate volume fraction ($\phi=0.26$) and cracks at large volume fraction ($\phi=0.40$). In both cases, the characteristic size, i.e., void diameter or crack width, measured after flow cessation corresponded to about 10 particle diameters. Subsequent studies by \cite{Tolpekin:2004,Conrad:2008} used faster CSLM to investigate the dynamic balance between aggregation and breakup events in depletion-induced gels of silica colloids under shear. More recently, the microstructure of colloidal gels made of poly(methyl-methacrylate) (PMMA) particles was imaged after imposing high-rate step strains, showing that bond-breaking events result from the erosion of rigid clusters that persist far beyond yielding and provide some degree of elasticity to the ruptured gel \cite{Hsiao:2012}.

In all these previous works, however, no time-resolved information on yielding could be extracted. To the best of our knowledge, the only time-resolved studies of yielding in colloidal gels under shear were performed by \cite{Rajaram:2010,Rajaram:2011}. There, fast CSLM allowed the authors to access the various stages of the shear-induced microstructure in depletion-induced gels of PMMA colloids. Although limited to a temporal resolution of a few seconds per confocal image, these experiments revealed that after a first regime where the structure turns from isotropic to oriented along the extensional component of the flow, yielding results from sporadic, local rupture events that sometimes lead to the formation of chain-like segments that eventually merge with the gel backbone as displayed in Fig.~\ref{fig:confocal}(b). The fully ruptured gel is constituted of dense, disconnected solid-like clusters that coexist with large voids, which is qualitatively consistent with the mesoscopic picture shown above in Fig.~\ref{fig:cbspatiotemporel}(b) for carbon black gels \cite{Gibaud:2010,Perge:2014b,Gibaud:2016}.

Finally, it appears that confocal microscopy has only been applied so far to colloidal gels that show reversible yielding. Exploring irreversible rupture in a time-resolved fashion at the particle or strand scale  constitutes one of the major challenges for future research. Experimental setups designed to apply a given shear stress while imaging the structure, such as those developed by \cite{Chan:2013,Dutta:2013,Aime:2016,Colombo:2019}, seem very promising to address this challenge, especially if coupled to a rheometer and if a spinning-disk confocal microscope is used in order to achieve even higher frame rates.

\section{Summary and conclusion}

Long-term goals in the study of yielding in colloidal gels are multiple. One of them consists in classifying the rupture processes in relation with the properties of the gels. This ``zoologist'' approach is all the more complex that the features of colloidal gels depend on a myriad of parameters such as the interaction potential, the particle concentration or the pathway taken to form the gel. Thus one issue is to determine the minimal level of description of the colloidal gels necessary to capture the yielding scenario. Here, we have described the current state of understanding of the nonlinear mechanical response of colloidal gels starting from experimental observations on two different systems that highlight the typical features of reversible fluidization and irreversible rupture. While carbon black gels display reversible fluidization preceded by strain softening and characterized by an exponential stress dependence of the yielding times, sodium caseinate gels show brittle rupture preceded by strain stiffening and failure times that depend on the load as power laws. Such a clear-cut distinction in the yielding behaviors prompts us to rationalize the nonlinear mechanics of colloidal gels based on their structure, in particular on their intermediate structures. Such an idea is supported by the fact that carbon black gels and sodium caseinate gels respectively display cluster- versus strand-based structures. This coarse-grained approach at the network level has the advantage to drastically diminish the number of parameters necessary to classify rupture, restricting it to cluster or strand properties, such as their size and their fractal dimension. In particular, it disregards the influence of parameters such as the colloid interactions, the colloid concentration and the gelation pathway. This could greatly simplify our understanding of rupture and set a simple basis for future theoretical and simulation work.

A more refined coarse-grained approach would consist in focusing on the particles. Such a ``colloid physicist'' approach has been followed in the early and current literature on colloid and protein dispersions and has allowed one to make fairly precise predictions about the phase behavior and the static properties of such particles under well-defined solution compositions \cite{cardinaux:2007,Gibaud:2009}. Coarse-grained approaches at the particle level are, however, far from providing a complete description of the complexity of colloidal gels. Colloidal particles are indeed rarely perfect spheres; their interaction potentials are in, general, not isotropic and do not necessarily remain invariant as a function of pH, temperature or concentration. For instance, carbon black particles are fractal aggregates that may penetrate one another at high concentrations. Sodium caseinate colloids have complex shape- and pH-dependent interactions that provide an original way to form gels starting from a fluid dispersion at equilibrium. Natural rubber latex gels, although very similar to standard latex particle gels, show unprecedented strain stiffening, which link with the reorganization of the microscopic network structure remains to be uncovered. Such spectacular properties, which make colloidal gels all the more versatile, are difficult to model and call for more refined approaches than simple coarse-graining, including complementary modeling based on polymer or macromolecular approaches. The limits of simple colloidal descriptions have been recently discussed in the case of globular proteins \cite{sarangapani:2015,Stradner:2020}. Similarly, disentangling the influence of colloid physics from the polymer or macromolecular aspects of the particle on the rupture properties of colloidal gels remains an open challenge and could be of great help to fine-tune the gel rupture behavior.

Therefore, beyond classifying yielding processes in colloidal gels, controlling and/or predicting their rupture appears as a major challenge for future work. There again, the adequate level of description is crucial. While mesoscopic measurements may help interpret macroscopic mechanical observations, direct visualization at the scale of individual strands or particles is essential to assess rupture precursors and localized events during yielding. We also emphasize that the discussions in the present paper may apply to a larger range of materials than just colloidal gels. Indeed, some features of shear-induced yielding reported here for colloidal gels share a number of similarities with the rupture of polymer gels \cite{Skrzeszewska:2010,Tabuteau:2009,Mora:2011,Erk:2012,Thornell:2014}, elastomers \cite{Ducrot2014} or hydrogels \cite{karobi2016,Bai:2019}, i.e., three-dimensional networks composed of macromolecules that are chemically or physically cross-linked. This suggests that either some specific details of the particle are irrelevant or that the ambivalent properties of the particles, half-way between colloids and polymer or macromolecules, are at the origin of this analogy. In any case, much work is still required before gel microstructures, nonlinear viscoelastic responses and yielding scenarios are fully connected and explained. Such a research effort will help to design and optimize soft materials where yielding is fully part of the application, e.g., in food products or 3D printing, or where it should rather be avoided.


\end{document}